\documentclass[aps,prd,reprint,nofootinbib,superscriptaddress,floatfix]{revtex4-2}

\usepackage{amsmath,amssymb,bm}
\usepackage{graphicx}
\usepackage{dcolumn}
\usepackage{booktabs}
\usepackage{hyperref}
\hypersetup{hidelinks}
\usepackage{microtype}
\usepackage{orcidlink}
\begin{document}

\title{Massive neutral Dirac quasibound states in a Newman--Janis-generated rotating charged Kalb--Ramond black-hole geometry}

\author{Sardor~Murodov\orcidlink{0000-0003-2360-4475}}
\email{s.murodov@newuu.uz}
\affiliation{New Uzbekistan University, Movarounnahr Street 1, Tashkent 100000, Uzbekistan}

\author{Olimjon Kholturayev\orcidlink{0009-0004-3758-9610}} 
\email{o.xoltorayev@newuu.uz}
\affiliation{New Uzbekistan University, Movarounnahr Street 1, Tashkent 100000, Uzbekistan}

\author{Bahodir~Ahmedov\orcidlink{0009-0006-7375-2731}}
\email{b.ahmedov@newuu.uz}
\affiliation{New Uzbekistan University, Movarounnahr Street 1, Tashkent 100000, Uzbekistan}

\author{Bekzod~Rahmatov\orcidlink{0009-0001-0394-650X}}
\email{rahmatovbekzod@samdu.uz}
\affiliation{New Uzbekistan University, Movarounnahr Street 1, Tashkent 100000, Uzbekistan}

\author{Islom~Egamberdiev\orcidlink{0009-0001-6156-9271}}
\email{egamberdiyev.islom@samdaqu.edu.uz}
\affiliation{Samarkand State Technical University named after Mirzo Ulugbek, Lolazor street 70, Samarqand 140143, Uzbekistan}

\author{Bobomurat~Ahmedov\orcidlink{0000-0002-1232-610X}}
\email{ahmedov@astrin.uz}
\affiliation{Institute of Theoretical Physics, National University of Uzbekistan, Tashkent 100174, Uzbekistan}
\affiliation{School of Physics, Harbin Institute of Technology, Harbin 150001, People’s Republic of China}

\date{\today}

\begin{abstract}
We study quasibound states of a massive neutral Dirac field in the Newman--Janis-generated rotating charged geometry used in Lorentz-violating Kalb--Ramond (KR) gravity. Rather than assuming that the Kerr separation survives the deformation, we first recast the metric in four-dimensional off-shell Carter form. The corresponding principal closed conformal Killing--Yano tensor then provides the hidden symmetry needed to separate the torsion-free, minimally coupled Dirac equation. We derive the horizon and large-radius boundary conditions, including the non-Minkowskian asymptotic normalization, and determine the complex spectrum from coupled angular and matrix radial continued fractions. The code reproduces the Kerr spectrum and is checked by truncation studies and independent two-sided radial integrations. In scans at fixed metric parameters, increasing the KR parameter moves most real frequencies toward the mass threshold and reduces most decay rates. These trends are not invariant, however, because the same variation changes both the asymptotic potential scale and the distance from extremality. A scan in which these two quantities are held fixed changes the absolute ground-state trend but preserves a reversal in the real-frequency ordering of the maximal-$m$, $j=3/2$, $\ell=1$ pair. The location of the crossing shifts with the normalization convention. Thus the level reordering is more robust than the individual monotonic trends in binding energy or lifetime within this background family.
\end{abstract}

\maketitle

\section{Introduction}
\label{sec:intro}

Massive fields can be gravitationally trapped outside a black-hole horizon and form quasibound states whose amplitudes slowly decay as flux is absorbed by the hole. The complex frequency therefore carries both the oscillation energy and the lifetime of the state. For a massive Dirac field on Kerr spacetime, the equations separate exactly and the spectrum can be obtained accurately with coupled angular and radial continued fractions~\cite{DolanDempsey2015}. Quasibound states of charged Dirac fields have also been studied in static charged backgrounds~\cite{HuangLiuZhai2017}, while the quasinormal spectrum of massive fermions in rotating spacetimes provides a related, but distinct, spectral problem~\cite{KonoplyaZhidenko2018}.

The separability of the spin-$1/2$ equation in rotating black-hole geometries is well established. Chandrasekhar obtained the massive Dirac separation in Kerr, Page extended it to Kerr--Newman, and Carter and McLenaghan identified the associated first-order symmetry operator~\cite{Chandrasekhar1976,Page1976,CarterMcLenaghan1979}. Later analyses developed the spectral theory, propagators, decay properties, and exact-function formulations for Kerr--Newman and related geometries~\cite{BaticSchmid2006,HeJing2006,FinsterEtAl2002,BelgiornoCacciatori2009,BelgiornoCacciatori2010,Kraniotis2019,DariescuEtAl2021,NeznamovEtAl2019}. Continued-fraction and bound-state calculations in charged and nonrotating backgrounds further established the numerical tools used for fermionic resonances~\cite{RichartzGiugno2014,Wu2009PRD,Wu2009CQG,Sporea2019,Dolan2007,FuruhashiNambu2004}.

For the present problem, the important point is that Kerr separability is tied to hidden geometry, not simply to the use of Boyer--Lindquist coordinates. Killing--Yano and conformal Killing--Yano tensors, together with the principal tensor of the canonical rotating class, generate the symmetry structure behind the separated equations~\cite{CarigliaKrtousKubiznak2011,FrolovKrtousKubiznak2018,FrolovKrtousKubiznak2017,FrolovKubiznak2007,HouriOotaYasui2009,Batista2015}. The corresponding symmetry operators and integrability conditions have been developed in a broad literature~\cite{KubiznakKunduriYasui2009,CayusoGrayKubiznak2019,BennCharltonKress1997,MasonTaghaviChabert2010,YasuiHouri2011,CarigliaEtAl2012,GibbonsRietdijkVanHolten1993,KalninsMillerWilliams1992,McLenaghanVanDenBergh1993,AhmedovAliev2009,LindstromSarioglu2022,WalkerPenrose1970}. This matters in a deformed rotating background: substituting a modified radial function into the Kerr equations is not enough by itself; one must first establish that the relevant hidden symmetry is still present.

Our numerical construction also follows methods developed for Kerr and Kerr--Newman perturbations. The Newman--Janis procedure provides a route from static to rotating metrics~\cite{Kerr1963,NewmanJanis1965,NewmanEtAl1965}, whereas Leaver's continued fractions remain a standard high-accuracy method for black-hole spectral problems~\cite{Teukolsky1973,Leaver1985,Leaver1990}. Related recurrence techniques have been used for massive fields, near-extremal modes, and high-precision resonance calculations in many settings~\cite{Cardoso2013,Rosa2010,Hod2016,HuangLiu2016,StottMarsh2018,KokkotasKonoplyaZhidenko2015,RosaDolan2012,BertiCardosoKokkotas2003,MarkYangZimmerman2015,ZimmermanMark2016,CookZalutskiy2014,Richartz2016,GajicWarnick2021}. Since the KR geometry considered here is obtained with a Newman--Janis-type construction, we treat it as a specified background and keep the known limitations of that procedure in modified gravity explicit~\cite{HansenYunes2013,Erbin2017,Shaikh2019}.

Kalb--Ramond fields are antisymmetric rank-two tensor fields that originally arose in string theory~\cite{KalbRamond1974}. When a KR field acquires a nonzero background value, nonminimal couplings can lead to Lorentz-violating black-hole geometries. Static and charged solutions and their thermodynamic, optical, and perturbative properties have been examined extensively in recent work~\cite{LessaEtAl2020,DuanZhaoYang2024,LiuWuWang2024,GuoTanLiu2024,LiuWuWei2025,TanLan2025,MangutSakalli2025,AraujoFilho2025}. Rotating or slowly rotating KR backgrounds have subsequently been used to study shadows, quasinormal modes, electromagnetic and particle dynamics, polarized images, EHT constraints, and rotational-energy extraction~\cite{LiuWuWang2025,ZahidEtAl2024,JumaniyozovEtAl2025,YangEtAl2026,YaoEtAl2026,NengrooIslamGhosh2026,HeEtAl2026}. Fermionic greybody factors and quasinormal modes have also been considered. Direct spinor couplings to KR-induced torsion, however, correspond to a different Dirac operator from the torsion-free probe adopted here~\cite{AlBadawiKraishan2023,PantigOvgunRincon2025,DeCesareMavromatosSarkar2015,SenGuptaSinha2001}.

Other strong-field applications of KR geometries include circular motion, QPO models, accretion radiation, PFDM environments, and plasma lensing~\cite{JumaniyozovEtAl2024,JumaniyozovEtAl2025QPO,JumaniyozovMurodovPFDM2025,RahmatovEgamberdievMurodov2025}. Related studies in deformed and regular black-hole spacetimes provide useful phenomenological comparisons~\cite{MurodovEtAlBBMB2025,RahmatovMurodovRayimbaev2025,MurodovRayimbaevAhmedov2023,AhmedMurodovRahmatov2026,RahmatovMurodovAhmedov2026,DonmezMurodovRayimbaev2026}. Here we focus on a different observable: the complex quasibound spectrum of a massive neutral spinor.

We are not aware of a previous calculation of massive neutral Dirac quasibound states for the rotating charged KR background studied below. Two features make the extension from Kerr nontrivial. First, separability has to be demonstrated for the deformed rotating metric rather than inherited by assumption. Second, the spacetime is not asymptotically Minkowski, so both the decaying large-radius solution and the normalization of the frequency differ from the Kerr case. These changes enter the boundary-value problem directly.

We address these points for a neutral test spinor coupled through the torsion-free Levi--Civita connection. The metric is first written in off-shell Carter form, which exposes its principal tensor and establishes the geometric basis for separation. We then derive the radial and angular systems, determine the near-horizon and large-radius exponents, and formulate a matrix continued fraction for the radial problem. The angular and radial conditions are solved simultaneously for the complex pair $(\omega,\lambda)$. Before turning on $l$ and $Q$, the implementation is tested against the Kerr results of Ref.~\cite{DolanDempsey2015}.

The hidden-symmetry argument should be distinguished from the KR-dependent part of the calculation. Separability follows from the off-shell Carter structure, whereas the KR model enters through the specific function $\Delta(r)$ and, consequently, through the horizon positions, the asymptotic falloff, and the radial recurrence. We examine the dependence of the spectrum on $l$, $Q$, $a$, and $M\mu$. Because varying $l$ also changes the asymptotic scale and the proximity to extremality, we supplement the fixed-parameter scan with a second scan in which those quantities are held fixed. This comparison turns out to be important when interpreting the frequency shifts.

The paper is organized as follows. Section~\ref{sec:geometry} describes the rotating charged KR geometry and its hidden symmetry. Section~\ref{sec:dirac-separation} gives the separated Dirac equations and the probe-field assumptions. The boundary conditions and continued-fraction construction are developed in Secs.~\ref{sec:boundary} and \ref{sec:continued-fraction}. Numerical results and consistency tests are presented in Sec.~\ref{sec:results}. Section~\ref{sec:conclusions} contains the discussion, and further analytical and numerical checks are collected in the Appendix.

\section{Rotating charged Kalb--Ramond geometry and hidden symmetry}
\label{sec:geometry}

We use the stationary and axisymmetric rotating charged KR geometry considered in Refs.~\cite{ZahidEtAl2024,JumaniyozovEtAl2025}. It is obtained within a Newman--Janis construction starting from a charged KR seed of the type discussed in Ref.~\cite{DuanZhaoYang2024}. In modified gravity such a construction does not, by itself, establish that the rotating metric satisfies the same field equations as the seed. We therefore take the line element below as the background of the spectral problem, without making an additional claim about its independent field-equation derivation. In Carter form,
\begin{align}
ds^2={}&-\frac{\Delta}{\Sigma}
\left(dt-a\sin^2\theta\,d\phi\right)^2
+\frac{\Sigma}{\Delta}\,dr^2+\Sigma\,d\theta^2
\nonumber\\
&+\frac{\sin^2\theta}{\Sigma}
\left[(r^2+a^2)d\phi-a\,dt\right]^2 ,
\label{eq:KRmetric-canonical}
\end{align}
where
\begin{align}
\Sigma&=r^2+a^2\cos^2\theta,\nonumber\\
\Delta(r)&=a^2-2Mr+\frac{r^2}{1-l}
+\frac{Q^2}{(1-l)^2}.
\label{eq:DeltaKR}
\end{align}
Here $M$, $Q$, and $a$ are metric parameters and $l$ is the dimensionless Lorentz-violating KR parameter. For $l\ne0$ the spacetime is not asymptotically Minkowski, so $M$ will not be identified a priori with an asymptotically flat ADM mass.

It is convenient to define
\begin{equation}
d\equiv 1-l>0,
\qquad
\Delta(r)=\frac{(r-r_+)(r-r_-)}{d},
\label{eq:defd}
\end{equation}
where the two roots are obtained directly from $\Delta=0$,
\begin{equation}
r_\pm
=
Md\pm\sqrt{M^2d^2-da^2-\frac{Q^2}{d}} .
\label{eq:horizons}
\end{equation}
Throughout the numerical analysis we restrict to the nonextremal
black-hole domain
\begin{equation}
M^2d^2-da^2-\frac{Q^2}{d}>0.
\label{eq:BHcondition}
\end{equation}

Equation~\eqref{eq:KRmetric-canonical} has the form of a four-dimensional off-shell Carter canonical metric~\cite{FrolovKrtousKubiznak2018}: the angular sector is unchanged from Kerr, while all model dependence enters through the radial function $\Delta(r)$. The principal-tensor construction of this canonical class does not require $\Delta(r)$ to be the Kerr polynomial or to satisfy the vacuum Einstein equations.

We introduce the orthonormal coframe
\begin{subequations}
\label{eq:Cartercoframe}
\begin{align}
e^{0}&=\sqrt{\frac{\Delta}{\Sigma}}
\left(dt-a\sin^2\theta\,d\phi\right),\\
e^{1}&=\sqrt{\frac{\Sigma}{\Delta}}\,dr,\\
e^{2}&=\sqrt{\Sigma}\,d\theta,\\
e^{3}&=\frac{\sin\theta}{\sqrt{\Sigma}}
\left[(r^2+a^2)d\phi-a\,dt\right].
\end{align}
\end{subequations}
Thus $ds^2=-(e^0)^2+(e^1)^2+(e^2)^2+(e^3)^2$.

\subsection{Principal tensor and Killing--Yano symmetry}
\label{subsec:principal}

Define the one-form
\begin{equation}
b=
-\frac12
\left[
\left(r^2+a^2\sin^2\theta\right)dt
-a\sin^2\theta\left(r^2+a^2\right)d\phi
\right].
\label{eq:principal-potential}
\end{equation}
Its exterior derivative,
\begin{equation}
h=db,
\label{eq:principal-h}
\end{equation}
can be written in the Carter frame as
\begin{equation}
h=r\,e^0\wedge e^1+a\cos\theta\,e^2\wedge e^3 .
\label{eq:principal-frame}
\end{equation}
No explicit form of $\Delta(r)$ appears in $h$. For $a\ne0$, the off-shell canonical construction therefore identifies $h$ as a nondegenerate closed conformal Killing--Yano two-form,
\begin{equation}
dh=0,
\qquad
\nabla_\gamma h_{\alpha\beta}
=
2g_{\gamma[\alpha}\xi_{\beta]} ,
\label{eq:CCKY}
\end{equation}
with, in four dimensions,
\begin{equation}
\xi_\beta=\frac{1}{3}\nabla^\alpha h_{\alpha\beta}.
\label{eq:principal-vector}
\end{equation}
The $a\to0$ limit is continuous, although the two-form becomes degenerate there. Its Hodge dual $f={}^\star h$ is a rank-two Killing--Yano tensor; its overall sign depends on the orientation convention and will not be needed below.

The principal tensor generates the usual hidden-symmetry tower of the canonical geometry~\cite{FrolovKrtousKubiznak2018}. In particular, it gives a first-order symmetry operator that commutes with the Dirac operator~\cite{CarigliaKrtousKubiznak2011}. The separation constant introduced below can therefore be viewed as an eigenvalue of this symmetry operator. This is the geometric reason that the Dirac equation remains separable after replacing the Kerr radial polynomial by the KR function $\Delta(r)$.

\section{Separation of the massive Dirac equation}
\label{sec:dirac-separation}

We consider a neutral test spinor of mass $\mu$. Its electromagnetic charge is set to zero, and the only coupling to the background is through the torsion-free Levi--Civita spin connection. No direct interaction with the electromagnetic potential or with a KR/torsion sector is included. These assumptions define the probe model used throughout the paper.

With metric signature $(-,+,+,+)$, we follow the gamma-matrix
convention of Ref.~\cite{DolanDempsey2015},
\begin{equation}
\{\gamma^\mu,\gamma^\nu\}=2g^{\mu\nu}\mathbf I_4 .
\label{eq:gamma-convention}
\end{equation}
In this convention the Dirac equation is written without an additional
overall factor of $i$,
\begin{equation}
\left(\gamma^\mu D_\mu-\mu\right)\Psi=0,
\qquad
D_\mu=\partial_\mu-\Gamma_\mu.
\label{eq:Dirac}
\end{equation}
The dual Carter tetrad is
\begin{subequations}
\begin{align}
e_0&=
\frac{r^2+a^2}{\sqrt{\Sigma\Delta}}\,\partial_t
+\frac{a}{\sqrt{\Sigma\Delta}}\,\partial_\phi,\\
e_1&=\sqrt{\frac{\Delta}{\Sigma}}\,\partial_r,\\
e_2&=\frac{1}{\sqrt{\Sigma}}\,\partial_\theta,\\
e_3&=
\frac{a\sin\theta}{\sqrt{\Sigma}}\,\partial_t
+\frac{1}{\sqrt{\Sigma}\sin\theta}\,\partial_\phi .
\end{align}
\end{subequations}

Introduce
\begin{equation}
\varrho=r+ia\cos\theta,
\qquad
\Sigma=\varrho\varrho^\ast.
\end{equation}
For the Carter tetrad, the radial and angular combinations of the contracted spin connection take the form
\begin{equation}
s_r=\sqrt{\frac{\Delta}{\Sigma}}
\left(\frac{1}{\varrho}+\frac{\Delta'}{2\Delta}\right),
\qquad
s_\theta=\frac{1}{\sqrt{\Sigma}}
\left(\cot\theta-\frac{ia\sin\theta}{\varrho}\right),
\label{eq:spinconnection-combinations}
\end{equation}
which is the same canonical structure as in Kerr~\cite{DolanDempsey2015}, with the present $\Delta(r)$ in the radial term. This suggests the rescaling

\begin{equation}
\Psi
=
\Delta^{-1/4}
\begin{pmatrix}
\varrho^{-1/2}\eta_-\\
(\varrho^\ast)^{-1/2}\eta_+
\end{pmatrix}.
\label{eq:spinor-rescaling}
\end{equation}
With
\begin{equation}
\eta_\pm(t,r,\theta,\phi)
=
e^{i(m\phi-\omega t)}
\eta_\pm(r,\theta),
\qquad
m\in\mathbb{Z}+\frac12,
\label{eq:harmonic}
\end{equation}
and
\begin{equation}
\eta_+
=
\begin{pmatrix}
R_1S_1\\
R_2S_2
\end{pmatrix},
\qquad
\eta_-
=
-
\begin{pmatrix}
R_2S_1\\
R_1S_2
\end{pmatrix},
\label{eq:product-ansatz}
\end{equation}
the Dirac equation separates into the canonical Kerr-type radial and angular systems~\cite{DolanDempsey2015},
\begin{subequations}
\label{eq:radial-separated}
\begin{align}
\sqrt{\Delta}
\left(\frac{d}{dr}-\frac{iK}{\Delta}\right)R_1
&=(\lambda+i\mu r)R_2,\\
\sqrt{\Delta}
\left(\frac{d}{dr}+\frac{iK}{\Delta}\right)R_2
&=(\lambda-i\mu r)R_1,
\end{align}
\end{subequations}
where
\begin{equation}
K(r)=(r^2+a^2)\omega-am.
\label{eq:Kdef}
\end{equation}
The angular operators are
\begin{equation}
\mathcal D_\pm
=\partial_\theta+\frac12\cot\theta
\pm m\csc\theta\mp a\omega\sin\theta.
\label{eq:angular-operators}
\end{equation}
The angular equations become
\begin{subequations}
\label{eq:angular-separated}
\begin{align}
\mathcal D_-S_1&=(\lambda+a\mu\cos\theta)S_2,\\
\mathcal D_+S_2&=(-\lambda+a\mu\cos\theta)S_1.
\end{align}
\end{subequations}
Thus $l$ and $Q$ enter the separated problem through the radial function $\Delta(r)$, whereas the angular equation is the familiar massive spin-$1/2$ spheroidal problem.

\section{Quasibound-state boundary conditions}
\label{sec:boundary}

Set
\begin{equation}
R_1=R_-,
\qquad
R_2=\sqrt{\Delta}\,R_+ .
\end{equation}
Then
\begin{subequations}
\begin{align}
\left(\frac{d}{dr}-\frac{iK}{\Delta}\right)R_-
&=(\lambda+i\mu r)R_+,\\
\left[
\frac{d}{dr}
+\frac{iK+\Delta'/2}{\Delta}
\right]R_+
&=
\frac{\lambda-i\mu r}{\Delta}R_- .
\end{align}
\end{subequations}

\subsection{Near-horizon behavior}

Define
\begin{equation}
\delta=r_+-r_-,
\qquad
K_+=(r_+^2+a^2)\omega-am.
\end{equation}
Near $r=r_+$,
\begin{equation}
\Delta
=
\frac{\delta}{d}(r-r_+)+O[(r-r_+)^2].
\end{equation}
Writing the leading horizon amplitudes as
\begin{equation}
R_-\sim A_H(r-r_+)^\sigma,
\qquad
R_+\sim B_H(r-r_+)^{\sigma-1},
\end{equation}
the ingoing Frobenius exponent is
\begin{equation}
\sigma=
\frac12-i\frac{dK_+}{\delta}.
\label{eq:sigmaKR}
\end{equation}
The leading terms of the first radial equation also fix the relative
horizon amplitude,
\begin{equation}
\left(\sigma-i\frac{dK_+}{\delta}\right)A_H
=(\lambda+i\mu r_+)B_H .
\label{eq:horizon-amplitude}
\end{equation}
Only one overall horizon normalization therefore remains free.
The horizon angular velocity with respect to the coordinate time $t$
and a convenient tortoise coordinate are
\begin{equation}
\Omega_H=\frac{a}{r_+^2+a^2},
\qquad
\frac{dr_*}{dr}=\frac{r^2+a^2}{\Delta}.
\end{equation}
Since $r_*\sim d(r_+^2+a^2)\ln(r-r_+)/\delta$ near $r_+$,
Eq.~\eqref{eq:sigmaKR} is the Frobenius form associated with an ingoing
phase proportional to $\exp[-i(\omega-m\Omega_H)r_*]$. Relative to the
normalized asymptotic time $T=t/\sqrt d$ introduced below, the horizon
angular velocity is
\begin{equation}
\Omega_{H,\mathrm{phys}}=\sqrt d\,\Omega_H,
\end{equation}
\begin{equation}
\omega_{\mathrm{phys}}-m\Omega_{H,\mathrm{phys}}
=\sqrt d\,(\omega-m\Omega_H).
\label{eq:physical-horizon-frequency}
\end{equation}

\subsection{Large-radius behavior}

For $r\to\infty$,
\begin{equation}
\Delta
=
\frac{r^2}{d}-2Mr+O(1).
\end{equation}
Using
\begin{equation}
R_-\sim r^\nu e^{qr},
\qquad
R_+\sim r^{\nu-1}e^{qr},
\end{equation}
the leading order gives
\begin{equation}
q=-\sqrt{d(\mu^2-d\omega^2)},
\qquad
\mathrm{Re}(q)<0,
\label{eq:qKR}
\end{equation}
whereas the next-to-leading order gives
\begin{equation}
\nu
=
\frac{Md^2(\mu^2-2d\omega^2)}{q}.
\label{eq:nuKR}
\end{equation}

The asymptotic normalization deserves some care. Since
\begin{equation}
g_{tt}\to-\frac1d,
\end{equation}
we define
\begin{equation}
T=\frac{t}{\sqrt d},
\qquad
\rho=\sqrt d\,r.
\end{equation}
The second definition normalizes the leading radial proper-distance element because $g_{rr}\to d$; it is not meant as a global proper-distance coordinate. The frequency conjugate to $T$ is
\begin{equation}
\omega_{\rm phys}=\sqrt d\,\omega.
\label{eq:physical-frequency}
\end{equation}
For a real bound-state frequency, the radial exponential becomes
\begin{equation}
e^{qr}
=
\exp\left[-\sqrt{\mu^2-\omega_{\rm phys}^2}\,\rho\right],
\label{eq:proper-radial-decay}
\end{equation}
so the usual threshold $|\omega_{\rm phys}|<\mu$ is recovered in these normalized variables. The quasibound frequencies are complex, however, and the practical boundary condition is the analytic choice $\mathrm{Re}(q)<0$, not an absolute-value inequality applied to complex $\omega$. Since $\sqrt{-g}=\Sigma\sin\theta$ grows only polynomially with radius, this branch also yields a convergent asymptotic Dirac norm for regular angular solutions.

This rescaling normalizes the asymptotic time and radial coordinate but does not make the spacetime asymptotically Minkowski. The angular sector still carries a solid-angle deformation, and rotation leaves a subleading $t$--$\phi$ cross term. The temporal component behaves as
\begin{equation}
g_{TT}=-1+\frac{2M d^{3/2}}{\rho}+O(\rho^{-2}),
\label{eq:gTT-asymptotic}
\end{equation}
which motivates the shorthand
\begin{equation}
\mathcal M_\infty\equiv M d^{3/2}.
\label{eq:Minfty}
\end{equation}
We use $\mathcal M_\infty$ as the coefficient of the leading asymptotic potential term, not as an ADM mass. For $Q=0$, the nonextremal condition is $a<M\sqrt d$, and it is useful to introduce
\begin{equation}
\chi_\Delta\equiv\frac{a}{M\sqrt d}<1.
\label{eq:chiDelta}
\end{equation}
This parameter measures the fraction of the $Q=0$ extremal rotation bound within the adopted metric family.

\section{Coupled angular and radial continued fractions}
\label{sec:continued-fraction}

\subsection{Angular continued fraction}

The angular equation is identical in form to the massive-Dirac angular problem in Kerr. Following Ref.~\cite{DolanDempsey2015}, let $P=\pm1$ be the parity label and define
\begin{equation}
\ell=j+\frac{P}{2},
\qquad
\epsilon_k=(-1)^{j-k}P,
\qquad
k=|m|,|m|+1,\ldots .
\end{equation}
The labels $(j,\ell,P)$ identify spectral branches by continuous continuation from the spherical/Kerr limits; for $a\ne0$ they should not be read as generators of an exact $SO(3)$ symmetry. Expanding the angular functions in spin-$1/2$ spherical harmonics gives
\begin{subequations}
\label{eq:angular-cf-coefficients}
\begin{align}
\alpha_k^{(\theta)}
&=(a\mu+\epsilon_k a\omega)
\frac{\sqrt{(k+1)^2-m^2}}{2(k+1)},\\
\beta_k^{(\theta)}
&=\epsilon_k\left(k+\frac12\right)
\left(1-\frac{am\omega}{k(k+1)}\right)
+\frac{a\mu m}{2k(k+1)}-\lambda,\\
\gamma_k^{(\theta)}
&=(a\mu-\epsilon_k a\omega)
\frac{\sqrt{k^2-m^2}}{2k}.
\end{align}
\end{subequations}
For the minimal solution we evaluate a backward continued fraction. At the largest retained index $k_{\max}$,
\begin{equation}
F_{k_{\max}}=\beta_{k_{\max}}^{(\theta)},
\end{equation}
and the downward recursion is
\begin{equation}
F_k=\beta_k^{(\theta)}
-\frac{\alpha_k^{(\theta)}\gamma_{k+1}^{(\theta)}}{F_{k+1}} .
\label{eq:angularCF}
\end{equation}
The angular eigenvalue condition is
\begin{equation}
F_{\rm ang}(\lambda,\omega)\equiv F_{|m|}=0.
\label{eq:angular-eigencondition}
\end{equation}
The symbol $N_{\rm ang}$ below denotes the corresponding finite angular truncation.

\subsection{Radial matrix continued fraction}

For the radial problem we adapt the matrix continued-fraction construction used for the Kerr Dirac equation in Ref.~\cite{DolanDempsey2015}. We write
\begin{equation}
\begin{aligned}
\begin{pmatrix}
R_-\\
\dfrac{r-r_+}{\sqrt d}R_+
\end{pmatrix}
={}&
\left(\frac{r-r_+}{\delta}\right)^\sigma
\left(\frac{r-r_-}{\delta}\right)^{-\sigma+\nu}
 e^{q(r-r_+)}\\
&\times\sum_{k=0}^{\infty}\boldsymbol{\xi}_k
\left(\frac{r-r_+}{r-r_-}\right)^k ,
\end{aligned}
\label{eq:KRseries}
\end{equation}
where
\[
\boldsymbol{\xi}_k\equiv
\begin{pmatrix}
\xi_k^{(-)}\\
\xi_k^{(+)}
\end{pmatrix}
\in\mathbb C^2 .
\]
The factors involving $\delta=r_+-r_-$ make the powers dimensionless. Relative to the equivalent prefactor $(r-r_+)^\sigma(r-r_-)^{-\sigma+\nu}e^{qr}$, they change only the overall radial normalization. The factor $(r-r_+)/\sqrt d$ in the second component is convenient because both entries then scale as $(r-r_+)^\sigma$ at the future horizon, while the large-$r$ behavior remains $R_-\sim r^\nu e^{qr}$ and $R_+\sim r^{\nu-1}e^{qr}$ up to constants.
With
\begin{equation}
x=\frac{r-r_+}{r-r_-},
\end{equation}
direct substitution into the first-order radial equations gives the three-term matrix recurrence
\begin{equation}
\boldsymbol{\alpha}_0\boldsymbol{\xi}_1
+\boldsymbol{\beta}_0\boldsymbol{\xi}_0=0,
\label{eq:matrix-recurrence-initial}
\end{equation}
and, for $k\ge1$,
\begin{equation}
\boldsymbol{\alpha}_k\boldsymbol{\xi}_{k+1}
+\boldsymbol{\beta}_k\boldsymbol{\xi}_k
+\boldsymbol{\gamma}_k\boldsymbol{\xi}_{k-1}=0.
\label{eq:matrix-recurrence}
\end{equation}
The exponent $\sigma$ already selects the ingoing Frobenius branch. The local horizon ratio in Eq.~\eqref{eq:horizon-amplitude} is therefore reserved for the independent shooting calculation and is not imposed as a second continued-fraction condition.
Let
\begin{equation}
K_\pm=(r_\pm^2+a^2)\omega-am,
\qquad
{\cal C}=(a^2+r_+r_-)\omega-am.
\end{equation}
Then
\begin{align}
\boldsymbol{\alpha}_k&=
\begin{pmatrix}
\alpha_{k1}&\alpha_{k2}\\0&\alpha_{k4}
\end{pmatrix},
&
\boldsymbol{\beta}_k&=
\begin{pmatrix}
\beta_{k1}&\beta_{k2}\\\beta_{k3}&\beta_{k4}
\end{pmatrix},\nonumber\\
\boldsymbol{\gamma}_k&=
\begin{pmatrix}
\gamma_{k1}&0\\\gamma_{k3}&\gamma_{k4}
\end{pmatrix}.
\end{align}
\begin{subequations}
\label{eq:KRrecurrence-coefficients}
\begin{align}
\alpha_{k1}&=k+\sigma+1-i\frac{dK_+}{\delta},\\
\alpha_{k2}&=-\sqrt d\,(\lambda+i\mu r_+),\\
\alpha_{k4}&=k+\sigma+\frac12+i\frac{dK_+}{\delta},\\
\beta_{k1}&=q\delta-2(k+\sigma)+\nu+2i\frac{d{\cal C}}{\delta},\\
\beta_{k2}&=\sqrt d\,(\lambda+i\mu r_-),\\
\beta_{k3}&=-\sqrt d\,(\lambda-i\mu r_+),\\
\beta_{k4}&=q\delta-2(k+\sigma)+\nu+1-2i\frac{d{\cal C}}{\delta},\\
\gamma_{k1}&=k+\sigma-\nu-1-i\frac{dK_-}{\delta},\\
\gamma_{k3}&=\sqrt d\,(\lambda-i\mu r_-),\\
\gamma_{k4}&=k+\sigma-\nu-\frac32+i\frac{dK_-}{\delta}.
\end{align}
\end{subequations}

For a minimal radial solution we introduce the tail matrix $\mathsf C_k$,
\begin{equation}
\mathsf C_k
=
-
\left(
\boldsymbol{\beta}_{k+1}
+\boldsymbol{\alpha}_{k+1}\mathsf C_{k+1}
\right)^{-1}
\boldsymbol{\gamma}_{k+1},
\label{eq:radial-tail}
\end{equation}
with the finite-cutoff initialization
\begin{equation}
\mathsf C_{N_{\rm rad}}=\mathbf 0 .
\label{eq:radial-tail-terminal}
\end{equation}
The radial eigenvalue condition is
\begin{equation}
\det\left(
\boldsymbol{\beta}_0+
\boldsymbol{\alpha}_0\mathsf C_0
\right)=0.
\label{eq:radialCF}
\end{equation}
A nonzero initial coefficient vector exists only when the matrix in Eq.~\eqref{eq:radialCF} is singular. In the Kerr limit these coefficients reduce to those of Ref.~\cite{DolanDempsey2015}. The shooting calculation, by contrast, starts from the local horizon relation \eqref{eq:horizon-amplitude}; agreement between the two methods therefore provides an independent check of the ingoing boundary condition.
The angular and radial equations together define
\begin{equation}
F_{\rm ang}(\lambda,\omega)=0,
\qquad
F_{\rm rad}(\lambda,\omega;M,a,Q,l,\mu)=0.
\end{equation}
We solve the real and imaginary parts of these two complex equations for $(\mathrm{Re}\,\omega,\mathrm{Im}\,\omega,\mathrm{Re}\,\lambda,\mathrm{Im}\,\lambda)$. Along parameter scans, the solution at one point is used as the initial guess for the next and both $\omega$ and $\lambda$ are followed continuously to limit branch switching. Convergence and independent radial checks are reported in Sec.~\ref{sec:results} and Appendix~\ref{app:numerical-audit}.

\section{Quasibound-state spectrum and Kalb--Ramond corrections}
\label{sec:results}

We now turn to the quasibound spectrum. Unless stated otherwise, the radial scale is set by $M=1$, with $M\mu=0.3$ and $a=0.9M$. Since the spacetime is not asymptotically Minkowski, $a/M$ and $Q/M$ should be understood here as ratios of metric parameters rather than automatically as asymptotic observables. Frequencies are quoted with respect to the normalized time coordinate $T=t/\sqrt{1-l}$,
\begin{equation}
\omega_{\rm phys}=\sqrt{1-l}\,\omega.
\end{equation}
For real frequencies, together with $\rho=\sqrt{1-l}\,r$, this gives the usual massive exponential threshold. For the complex modes, the decaying branch is selected directly by $\mathrm{Re}(q)<0$.

As a benchmark, setting $l=Q=0$ must recover the Kerr problem. For the eight modes used in the comparison, our results agree with Dolan and Dempsey~\cite{DolanDempsey2015} to better than $5\times10^{-9}$ in $\mathrm{Re}(\omega/\mu)$ and $4\times10^{-10}$ in $\mathrm{Im}(\omega/\mu)$. This test probes the angular and radial continued fractions simultaneously.

\begin{table}[tbp]
\caption{Representative Kerr-limit validation at $M\mu=0.3$ and $a=0.9M$. The last column gives the absolute difference in $\mathrm{Re}(\omega/\mu)$ from Ref.~\cite{DolanDempsey2015}.}
\label{tab:kerr-validation}
\begin{ruledtabular}
\begin{tabular}{lccc}
Mode & $\mathrm{Re}(\omega/\mu)$ & $\mathrm{Im}(\omega/\mu)$ & $|\Delta\mathrm{Re}|$ \\
$j=1/2,\,\ell=0,\,m=-0.5$ & 0.951205798 & -3.129e-02 & 2.0e-09 \\
$j=1/2,\,\ell=0,\,m=+0.5$ & 0.922000865 & -2.319e-02 & 4.9e-09 \\
$j=1/2,\,\ell=1,\,m=+0.5$ & 0.986626308 & -6.306e-05 & 2.0e-09 \\
$j=3/2,\,\ell=1,\,m=+1.5$ & 0.987988499 & -4.200e-08 & 1.4e-09 \\
\end{tabular}
\end{ruledtabular}
\end{table}

\begin{figure}[tbp]
\centering
\includegraphics[width=0.80\columnwidth]{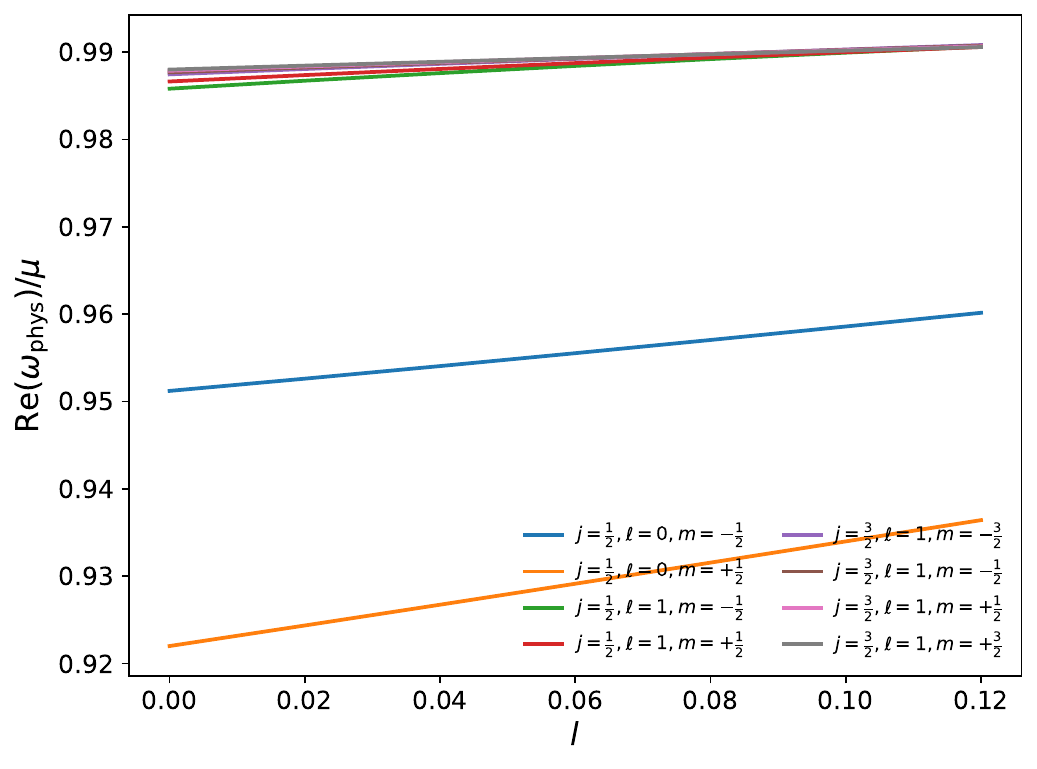}
\caption{Real part of the asymptotically normalized quasibound-state frequency in the fixed-metric-parameter scan, $M\mu=0.3$, $a=0.9M$, and $Q=0$. Equation~\eqref{eq:default-scan-drift} shows that this convention does not hold the asymptotic potential scale or the fraction of extremality fixed. The larger-$l$ part is exploratory.}
\label{fig:omega-l}
\end{figure}

\begin{figure}[tbp]
\centering
\includegraphics[width=0.80\columnwidth]{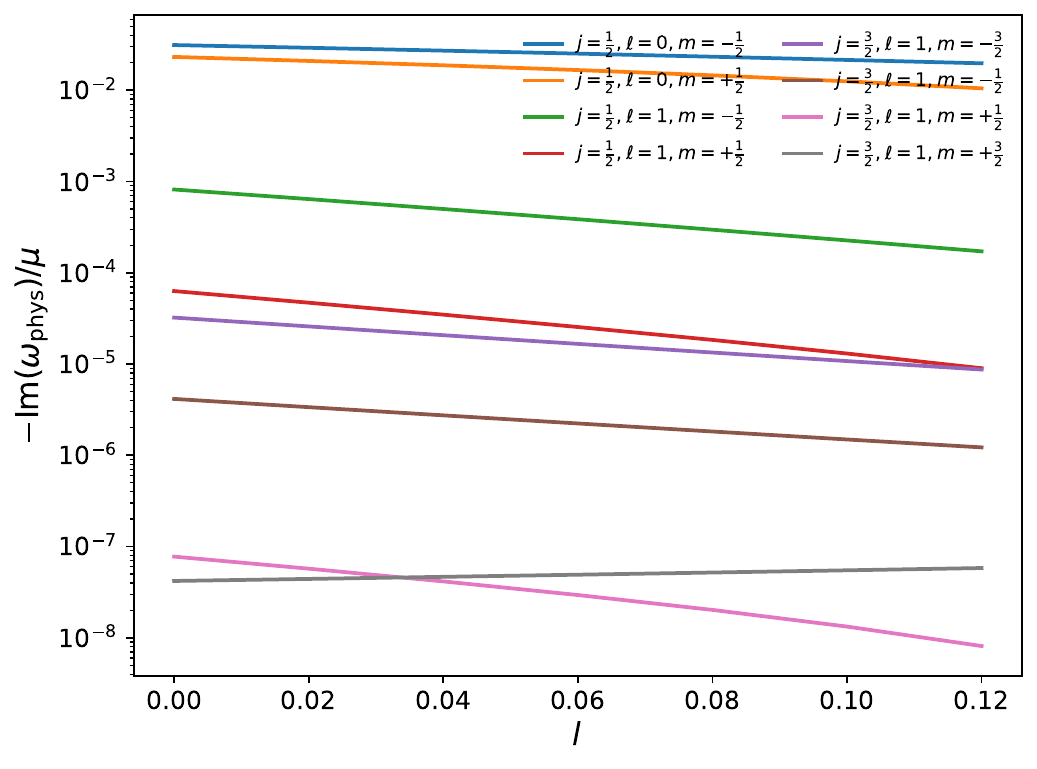}
\caption{Decay rate $-\mathrm{Im}(\omega_{\rm phys})/\mu$ for the same fixed-metric-parameter scan as in Fig.~\ref{fig:omega-l}.}
\label{fig:im-l}
\end{figure}

We first set $Q=0$ and vary $l$ while keeping $M=1$ and $a/M=0.9$ fixed. In this family, the normalized real frequencies of the displayed modes generally move upward toward the mass threshold as $l$ increases. Seven of the eight modes also show smaller decay rates. The maximally corotating $j=3/2$, $\ell=1$, $m=+3/2$ state is the exception: its already small decay rate increases slightly. Figures~\ref{fig:omega-l} and \ref{fig:im-l} show these trends. They describe the chosen fixed-metric-parameter family, but they should not yet be interpreted as invariant statements about how the physical binding scale changes with $l$.

The source of this ambiguity is explicit in Eqs.~\eqref{eq:Minfty} and \eqref{eq:chiDelta}. At fixed $M\mu=0.3$ and $a/M=0.9$,
\begin{equation}
\mathcal M_\infty\mu=0.3(1-l)^{3/2},
\qquad
\chi_\Delta=\frac{0.9}{\sqrt{1-l}},
\label{eq:default-scan-drift}
\end{equation}
so a change in $l$ simultaneously changes the asymptotic potential scale and the fraction of the $Q=0$ extremal rotation bound. To separate these effects, we repeat the scan with $\mathcal M_\infty\mu=0.3$ and $\chi_\Delta=0.9$ fixed. Taking $\mathcal M_\infty=1$ as the length unit along this family gives
\begin{equation}
M=d^{-3/2},
\qquad
\mu=0.3,
\qquad
a=\frac{0.9}{d}.
\label{eq:controlled-family}
\end{equation}
For compact notation in Table~\ref{tab:normalization-control}, define
\begin{subequations}
\label{eq:controlled-definitions}
\begin{align}
\omega_{\rm g}
&\equiv \omega_{j=1/2,\,\ell=0,\,m=+1/2},\nonumber\\
\omega_\pm^{(3/2)}
&\equiv \omega_{j=3/2,\,\ell=1,\,m=\pm3/2},\nonumber\\
\Delta_m
&\equiv\frac{\mathrm{Re}(\omega_{+,\rm phys}^{(3/2)}-
\omega_{-,\rm phys}^{(3/2)})}{\mu}.
\end{align}
\end{subequations}
The controlled scan reverses the absolute ground-state trend: $\mathrm{Re}(\omega_{\rm phys})/\mu$ decreases and the decay rate increases with $l$ for the $j=1/2$, $\ell=0$, $m=+1/2$ mode. The monotonic behavior seen in the fixed-$M$ scan is therefore not independent of the parameter normalization.

\begin{table}[tbp]
\caption{Normalization-controlled $Q=0$ scan with
$\mathcal M_\infty\mu=0.3$ and $\chi_\Delta=0.9$. The last column is the
maximal-$m$ $j=3/2$, $\ell=1$ real-frequency splitting.}
\label{tab:normalization-control}
\begin{ruledtabular}
\begin{tabular}{cccc}
$l$ & $\mathrm{Re}(\omega_{{\rm g},{\rm phys}})/\mu$ &
$-\mathrm{Im}(\omega_{{\rm g},{\rm phys}})/\mu$ & $\Delta_m$\\
0.00 & 0.922000865 & 0.023193867 & +5.455e-04\\
0.04 & 0.916903790 & 0.029735399 & +3.905e-04\\
0.08 & 0.911690540 & 0.037864457 & +1.928e-04\\
0.12 & 0.906433592 & 0.047990829 & -6.078e-05\\
\end{tabular}
\end{ruledtabular}
\end{table}

\begin{figure}[tbp]
\centering
\includegraphics[width=0.80\columnwidth]{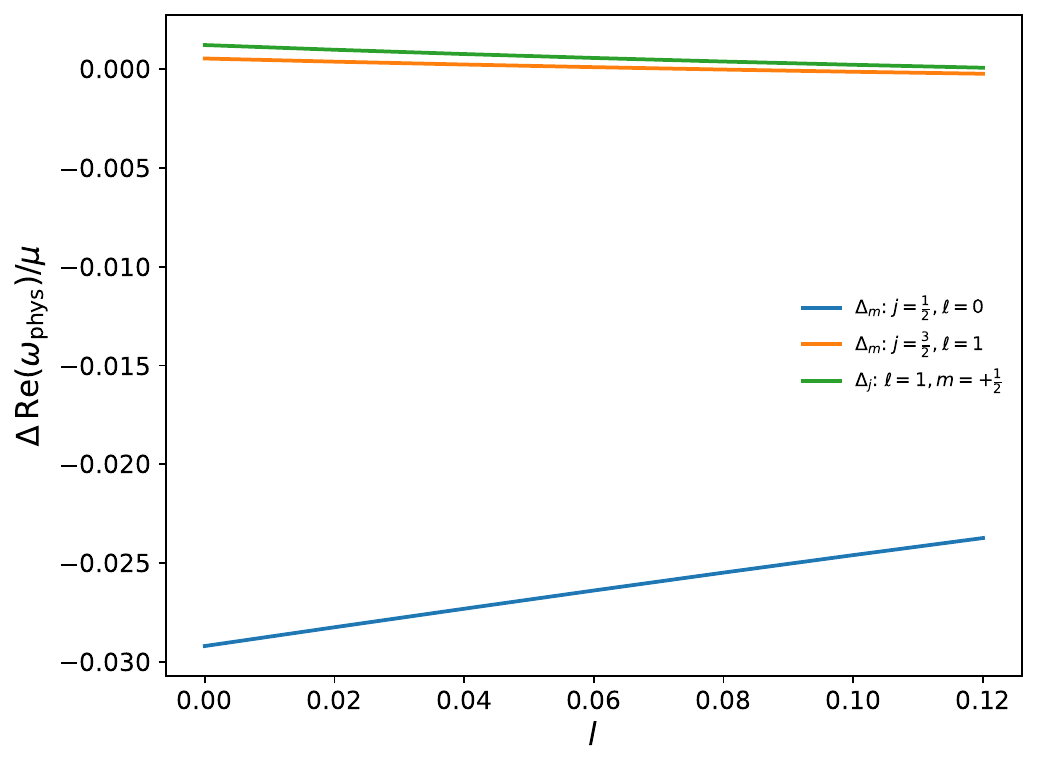}
\caption{Selected rotational and $j$-dependent real-frequency splittings in the fixed-$M$ scan. Fine continuation locates the maximal-$m$ $j=3/2$, $\ell=1$ sign reversal at $l_\star\simeq0.07747$. The normalization-controlled scan shifts this zero to $l_\star\simeq0.11129$.}
\label{fig:splittings}
\end{figure}

The level ordering of the maximal-$m$, $j=3/2$, $\ell=1$ pair behaves differently. In the fixed-$M$ scan, continuation locates a zero of the real-frequency splitting at
\begin{equation}
l_\star^{(M)}\simeq0.07747,
\end{equation}
whereas the controlled family gives
\begin{equation}
l_\star^{(\mathcal M_\infty,\chi)}\simeq0.11129.
\end{equation}
Only the real parts cross; the complex eigenfrequencies remain distinct. At the fixed-$M$ zero, using the notation of Eq.~\eqref{eq:controlled-definitions},
\begin{align}
\omega_{-,{\rm phys}}^{(3/2)}/\mu
&=0.989693201979-1.37473753\times10^{-5}i,\\
\omega_{+,{\rm phys}}^{(3/2)}/\mu
&=0.989693201855-5.15955330\times10^{-8}i.
\end{align}
Thus the sign reversal itself survives the normalization control, even though the value of $l$ at which it occurs does not. Figure~\ref{fig:splittings} displays this reordering for the fixed-$M$ family.

\begin{figure}[tbp]
\centering
\includegraphics[width=0.80\columnwidth]{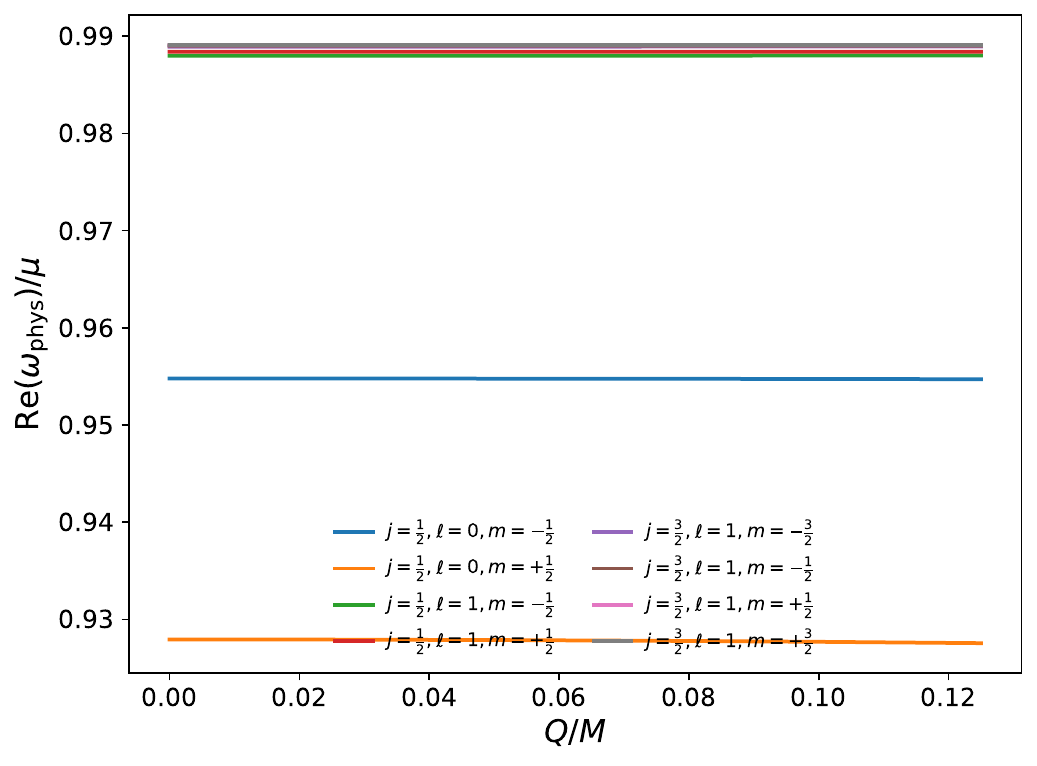}
\caption{Real part of the asymptotically normalized quasibound-state frequency as a function of the metric-parameter ratio $Q/M$ at $l=0.05$, $M\mu=0.3$, and $a=0.9M$.}
\label{fig:omega-q}
\end{figure}

At fixed $l=0.05$, the charge parameter has a much smaller effect on the real frequencies; see Fig.~\ref{fig:omega-q}. For the $j=1/2$, $\ell=0$, $m=+1/2$ state, changing $Q/M$ from $0$ to $0.125$ shifts $\mathrm{Re}(\omega_{\rm phys})/\mu$ by about $-3.8\times10^{-4}$, while the higher-$\ell$ branches move only at the $10^{-5}$ level. This weak dependence is natural for the neutral probe because $Q$ affects the Dirac equation only through the metric. Moreover, Eq.~\eqref{eq:DeltaKR} contains $Q$ only as $Q^2$. The neutral spectrum is therefore invariant under $Q\to-Q$, and its leading small-$Q$ correction is quadratic.

\begin{figure}[tbp]
\centering
\includegraphics[width=0.80\columnwidth]{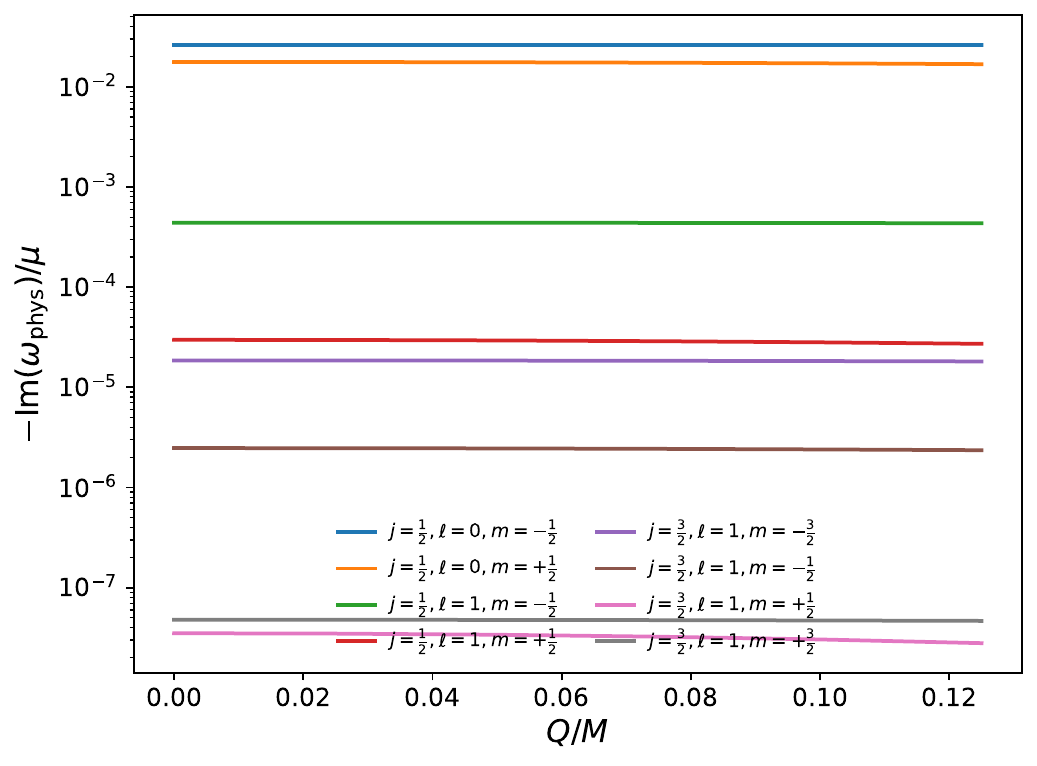}
\caption{Decay rate $-\mathrm{Im}(\omega_{\rm phys})/\mu$ as a function of $Q/M$ for the same parameters as in Fig.~\ref{fig:omega-q}. The weak variation reflects the indirect, metric-only charge dependence of the neutral probe.}
\label{fig:im-q}
\end{figure}

The decay rates show the same weak charge sensitivity in Fig.~\ref{fig:im-q}. There is no direct electromagnetic interaction in the present probe model, so the charge can influence a mode only by changing the background geometry. Since that geometric dependence is even in $Q$, both the real and imaginary parts of the spectrum have vanishing linear corrections at $Q=0$.

\begin{figure}[tbp]
\centering
\includegraphics[width=0.80\columnwidth]{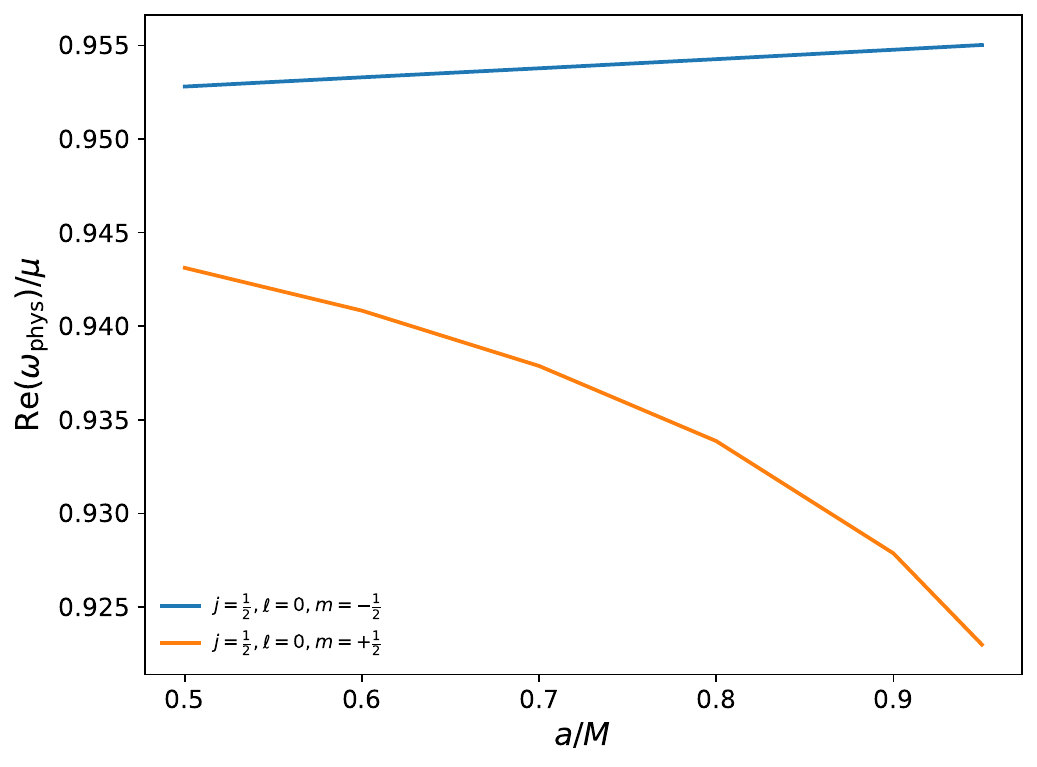}
\caption{Spin dependence of the $j=1/2$, $\ell=0$ doublet for $l=0.05$, $Q=0.05M$, and $M\mu=0.3$. Increasing $a/M$ progressively separates the $m=\pm1/2$ branches.}
\label{fig:spin}
\end{figure}

The spin dependence is illustrated in Fig.~\ref{fig:spin}. As $a/M$ is increased, the $j=1/2$, $\ell=0$, $m=+1/2$ branch moves to lower $\mathrm{Re}(\omega_{\rm phys})/\mu$, whereas the $m=-1/2$ branch changes only slightly in the opposite direction. Rotation therefore separates the two azimuthal branches rather than shifting them together. With our sign convention, the lower real frequency corresponds to a larger deficit from the mass threshold, so the $m=+1/2$ state becomes more strongly bound over the range shown. This is the same type of rotational splitting familiar from the Kerr massive-Dirac spectrum, now evaluated in the deformed background.

\begin{figure}[tbp]
\centering
\includegraphics[width=0.80\columnwidth]{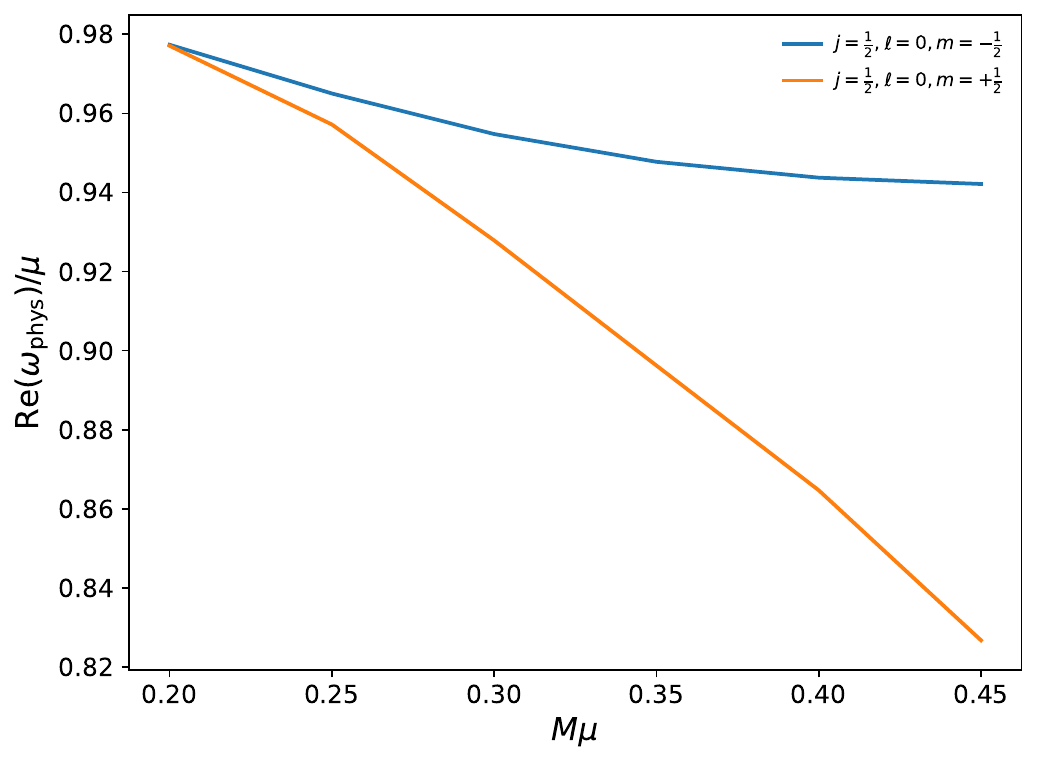}
\caption{Dependence of the $j=1/2$, $\ell=0$ doublet on the dimensionless mass parameter $M\mu$ for $a=0.9M$, $l=0.05$, and $Q=0.05M$. Both branches move farther below the mass threshold as $M\mu$ increases, with a stronger shift for $m=+1/2$.}
\label{fig:coupling}
\end{figure}

Figure~\ref{fig:coupling} shows the dependence on $M\mu$. Increasing the field--black-hole coupling lowers $\mathrm{Re}(\omega_{\rm phys})/\mu$ for both members of the $j=1/2$, $\ell=0$ doublet, so the modes lie farther below the mass threshold. The change is larger for $m=+1/2$, and the rotational splitting consequently grows across the plotted interval. We regard this as a finite-range numerical trend of the present background rather than an asymptotic statement for arbitrarily large $M\mu$.

For the final survey, every stored parameter--mode combination was recomputed at the same truncation, $(N_{\rm ang},N_{\rm rad})=(260,440)$. The largest angular residual is $5.2\times10^{-14}$ and the largest radial residual is $2.51\times10^{-11}$ over the 136 roots; all modes have $\mathrm{Im}(\omega_{\rm phys})<0$. Repeating the full survey at $(220,360)$ changes $\omega/\mu$ by at most $6.85\times10^{-9}$. Root acceptance is based on these residual and truncation checks rather than on the nonlinear solver status alone.

For the representative mode $l=0.10$, $Q=0.05M$, $j=1/2$, $\ell=0$, $m=+1/2$, the continued-fraction root is stable through several further digits as the truncation is increased; the sequence is listed in Table~\ref{tab:cf-convergence}. We also checked selected deformed modes by integrating a Riccati form of the radial equations from both boundaries and matching at an intermediate radius. For three $j=1/2$, $\ell=0$ cases with nonzero deformation and charge, the relative mismatch lies between $7\times10^{-9}$ and $2\times10^{-7}$.

Because the level crossing is the most distinctive feature of the spectrum, the same two-sided integration was applied directly to both $m=\pm3/2$ modes at each crossing. We integrate $w=(r-r_+)R_+/(\sqrt d\,R_-)$ from the horizon using Eq.~\eqref{eq:horizon-amplitude} and from large radius using a fifth-order expansion of the decaying solution. For all four crossing modes the relative matching mismatch is below $1.5\times10^{-7}$; details are given in Appendix~\ref{app:crossing-shooting}. This test is independent of the radial continued-fraction eigencondition.

\section{Discussion and conclusions}
\label{sec:conclusions}

We have studied massive neutral Dirac quasibound states in the adopted Newman--Janis-generated rotating charged KR background. The first point that has to be settled is separability. Written in off-shell Carter form, the metric admits the principal closed conformal Killing--Yano tensor, so the torsion-free minimally coupled Dirac equation separates for the same geometric reason as in the broader canonical class. The KR dependence then enters through the radial function $\Delta(r)$, which changes the horizon structure, the asymptotic behavior, and the radial spectral equation.

The large-radius behavior requires a separate normalization because the geometry is not asymptotically Minkowski. With $T=t/\sqrt{1-l}$ and $\rho=\sqrt{1-l}\,r$, the frequency measured with respect to the normalized asymptotic time is $\omega_{\rm phys}=\sqrt{1-l}\,\omega$, while the decaying complex branch is fixed by $\mathrm{Re}(q)<0$. The same expansion gives $\mathcal M_\infty=M(1-l)^{3/2}$ as the coefficient of the leading $1/\rho$ term in $g_{TT}$. We use this quantity only as an asymptotic normalization scale; the remaining solid-angle deformation prevents an ordinary ADM interpretation.

This point is important for the interpretation of the numerical scans. If $M$, $a/M$, and $M\mu$ are held fixed while $l$ is increased, most normalized real frequencies move toward the mass threshold and most decay rates decrease. At the same time, however, $\mathcal M_\infty\mu$ decreases and the $Q=0$ background approaches extremality. In a second scan with $\mathcal M_\infty\mu=0.3$ and $\chi_\Delta=a/[M\sqrt{1-l}]=0.9$ fixed, the ground-state frequency and decay rate move in the opposite direction. The absolute monotonic trends therefore depend on what is held fixed when $l$ is varied.

A more stable feature is the reordering of the maximal-$m$, $j=3/2$, $\ell=1$ pair. In the fixed-$M$ scan at $M\mu=0.3$, $a=0.9M$, and $Q=0$, the real-frequency splitting changes sign at $l_\star\simeq0.07747$. The imaginary parts remain different, so this is not a degeneracy of the full complex modes. The same ordering reversal is present in the normalization-controlled scan, where the zero moves to $l_\star\simeq0.11129$. The existence of the reversal is therefore less sensitive to the normalization convention than its numerical location. For the neutral probe, the charge dependence is even in $Q$ because the metric contains $Q$ only through $Q^2$, so the leading small-charge shift is quadratic.

Several checks were used to test the numerical implementation. The code reproduces the published Kerr frequencies, the radial recurrence can be reconstructed directly from the transformed first-order system, and all 136 survey roots were recomputed at a common truncation. Selected deformed modes were also checked by two-sided radial integration, including both members of the level-crossing pair. At $(N_{\rm ang},N_{\rm rad})=(260,440)$ the largest angular and radial residuals are $5.2\times10^{-14}$ and $2.51\times10^{-11}$, respectively. Changing the common truncation from $(220,360)$ to $(260,440)$ alters $\omega/\mu$ by no more than $6.85\times10^{-9}$ over the stored survey, and the dedicated crossing integrations give relative matching mismatches below $1.5\times10^{-7}$.

The calculation is deliberately restricted to a neutral metric probe. The spinor has no direct electromagnetic coupling and no explicit coupling to the KR two-form or to a torsionful spin connection. Charged or torsion-coupled fermions would require a different Dirac operator and, in general, a new separation analysis. Within the assumptions used here, the main result is therefore the persistence of a KR-induced rotational level reordering, while statements about absolute binding energies and lifetimes must always be tied to a specified parameter normalization.

\section*{Acknowledgments}
S.M. gratefully acknowledges support from Grant FZ-20200929385 of the Agency of Innovative Developments of the Republic of Uzbekistan.

\section*{Data Availability}
The numerical tables underlying the figures and the custom solver and audit scripts are available from the corresponding author upon reasonable request.

\appendix
\section{Analytical and numerical audit}
\label{app:checks}

\subsection{Principal two-form}

From Eq.~\eqref{eq:principal-potential},
\begin{align}
db={}&
-r\,dr\wedge dt
+a r\sin^2\theta\,dr\wedge d\phi
\nonumber\\
&-a^2\sin\theta\cos\theta\,d\theta\wedge dt
\nonumber\\
&+a(r^2+a^2)\sin\theta\cos\theta
\,d\theta\wedge d\phi .
\end{align}
Using Eq.~\eqref{eq:Cartercoframe}, this is
\begin{equation}
r\,e^0\wedge e^1+a\cos\theta\,e^2\wedge e^3.
\end{equation}
Thus the two-form itself contains no explicit dependence on $\Delta(r)$ and $dh=d^2b=0$ identically. The conformal Killing--Yano equation follows from the general off-shell Carter canonical construction~\cite{FrolovKrtousKubiznak2018}.

\subsection{Spin-connection reduction}

The Carter-frame connection combinations entering the separated Dirac operator can be written as
\begin{align}
s_r&=e_1{}^r\frac{1}{\varrho\sqrt{\Delta}}
\frac{\partial}{\partial r}\left(\varrho\sqrt{\Delta}\right)
=\sqrt{\frac{\Delta}{\Sigma}}
\left(\frac1\varrho+\frac{\Delta'}{2\Delta}\right),\\
s_\theta&=e_2{}^\theta\frac{1}{\varrho\sin\theta}
\frac{\partial}{\partial\theta}\left(\varrho\sin\theta\right)
=\frac1{\sqrt\Sigma}
\left(\cot\theta-\frac{ia\sin\theta}{\varrho}\right).
\end{align}
The rescaling in Eq.~\eqref{eq:spinor-rescaling} removes these connection factors from the first-order derivative terms, leaving the canonical separated system in Eqs.~\eqref{eq:radial-separated} and \eqref{eq:angular-separated}. In the Kerr limit these expressions coincide with the Carter-tetrad construction of Ref.~\cite{DolanDempsey2015}.

\subsection{Radial-recurrence reconstruction}

The radial recurrence was also reconstructed directly from the first-order differential system. With $x=(r-r_+)/(r-r_-)$ and the prefactor in Eq.~\eqref{eq:KRseries}, collecting equal powers of $x$ gives Eqs.~\eqref{eq:matrix-recurrence-initial} and \eqref{eq:matrix-recurrence}, including the coefficients in Eq.~\eqref{eq:KRrecurrence-coefficients}. This is the same matrix three-term organization used for the Kerr Dirac problem in Ref.~\cite{DolanDempsey2015}. A separate Frobenius expansion at $r=r_+$ gives Eq.~\eqref{eq:horizon-amplitude}; this local ratio is used only to initialize the ODE shooting calculation and is not an additional continued-fraction condition.

\subsection{Kerr limit}

For $l=0$ and $Q=0$,
\begin{equation}
d=1,\qquad
\Delta=r^2-2Mr+a^2,\qquad
r_+r_-=a^2.
\end{equation}
Equations~\eqref{eq:qKR} and \eqref{eq:nuKR} reduce to
\begin{equation}
q=-\sqrt{\mu^2-\omega^2},
\qquad
\nu=\frac{M(\mu^2-2\omega^2)}{q}.
\end{equation}
Moreover,
\begin{equation}
K_\pm=2M\omega r_\pm-am,
\qquad
{\cal C}=2a^2\omega-am.
\end{equation}
Multiplying the matrices in Eq.~\eqref{eq:KRrecurrence-coefficients} by the common factor $r_+-r_-$ recovers the Kerr recurrence coefficients used by Dolan and Dempsey~\cite{DolanDempsey2015}. The numerical implementation is anchored by reproducing their tabulated Kerr modes before $l$ and $Q$ are switched on.

\subsection{Continued-fraction convergence and crossing refinement}
\label{app:numerical-audit}

Table~\ref{tab:cf-convergence} gives the truncation sequence for the representative deformed mode $M=1$, $M\mu=0.3$, $a=0.9M$, $l=0.10$, $Q=0.05M$, $j=1/2$, $\ell=0$, $m=+1/2$. The table lists the coordinate frequency used internally by the recurrence; the frequencies plotted in the main text are multiplied by $\sqrt{1-l}$.

\begin{table}[tbp]
\caption{Continued-fraction truncation convergence for a representative deformed mode.}
\label{tab:cf-convergence}
\begin{ruledtabular}
\begin{tabular}{cccc}
$N_{\rm ang}$ & $N_{\rm rad}$ & $\mathrm{Re}(\omega/\mu)$ & $-\mathrm{Im}(\omega/\mu)$\\
100 & 160 & 0.984436859370 & 0.013016546104\\
140 & 220 & 0.984436862567 & 0.013016558992\\
180 & 280 & 0.984436862896 & 0.013016560617\\
220 & 360 & 0.984436862945 & 0.013016560926\\
260 & 440 & 0.984436862950 & 0.013016560968\\
\end{tabular}
\end{ruledtabular}
\end{table}

For the maximal-$m$, $j=3/2$, $\ell=1$ splitting at $Q=0$, high-truncation continuation in the fixed-$M$ family gives
\begin{equation}
l_\star^{(M)}\simeq0.07747.
\end{equation}
Writing $\omega_\pm\equiv\omega_{m=\pm3/2}$, the two frequencies at this point are
\begin{align}
\omega_-^{\rm phys}/\mu
&=0.989693201979-1.37473753\times10^{-5}i,\\
\omega_+^{\rm phys}/\mu
&=0.989693201855-5.15955330\times10^{-8}i.
\end{align}
The real parts coincide to the displayed accuracy, whereas the decay rates remain different.

Repeating the continuation at fixed $\mathcal M_\infty\mu=0.3$ and $\chi_\Delta=0.9$ gives
\begin{equation}
l_\star^{(\mathcal M_\infty,\chi)}\simeq0.11129,
\end{equation}
with
\begin{align}
\omega_-^{\rm phys}/\mu
&=0.985940261980-8.53895246\times10^{-5}i,\\
\omega_+^{\rm phys}/\mu
&=0.985940262192-1.81283256\times10^{-7}i.
\end{align}
The reversal therefore survives the change of normalization, although the crossing position shifts. The unrounded continuation roots leave real-frequency splitting residuals of order $10^{-10}$; these numbers are diagnostics of the root finder, not uncertainties in $l_\star$. Since a separate cutoff sequence for the crossing position was not computed, we quote $l_\star$ to five decimal places.

\subsection{Independent shooting at the level crossings}
\label{app:crossing-shooting}

As an independent test of the level crossings, define
\begin{equation}
w(r)=\frac{r-r_+}{\sqrt d}\frac{R_+(r)}{R_-(r)}.
\end{equation}
The first-order radial equations give a Riccati equation for $w$. We integrate outward from $r=r_++10^{-6}$ using the ingoing ratio in Eq.~\eqref{eq:horizon-amplitude}, and inward from $r_{\rm out}=400$ using a fifth-order $1/r$ expansion of the decaying solution. The two integrations are matched at $r_{\rm m}=6$, with
\begin{equation}
\epsilon_{\rm shoot}=\frac{|w_{\rm H}(r_{\rm m})-w_{\infty}(r_{\rm m})|}
{\max\{1,|w_{\rm H}(r_{\rm m})|,|w_{\infty}(r_{\rm m})|\}}.
\end{equation}
The angular separation constant is obtained independently from the angular continued fraction at each fixed complex frequency. The mismatches at $r_{\rm out}=400$ are listed in Table~\ref{tab:crossing-shooting}.
\begin{table}[tbp]
\caption{Independent two-sided radial shooting check for both maximal-$m$
modes at the two reported real-frequency crossings.}
\label{tab:crossing-shooting}
\begin{ruledtabular}
\begin{tabular}{lccc}
Family & $m$ & $l_\star$ & $\epsilon_{\rm shoot}$\\
fixed $M$ & $-3/2$ & 0.07747 & $1.20\times10^{-7}$\\
fixed $M$ & $+3/2$ & 0.07747 & $1.45\times10^{-7}$\\
fixed $\mathcal M_\infty,\chi_\Delta$ & $-3/2$ & 0.11129 & $5.51\times10^{-8}$\\
fixed $\mathcal M_\infty,\chi_\Delta$ & $+3/2$ & 0.11129 & $1.43\times10^{-7}$\\
\end{tabular}
\end{ruledtabular}
\end{table}
Varying the outer boundary from $r_{\rm out}=250$ to 600 leaves the mismatches unchanged at the displayed precision.

\subsection{Common-truncation survey audit}
\label{app:survey-residual}

All 136 stored parameter--mode combinations were re-solved with the common truncation
\begin{equation}
(N_{\rm ang},N_{\rm rad})=(260,440)
\end{equation}
and nonlinear-solver parameter tolerance $10^{-11}$. The roots were then assessed from their residuals and truncation stability. The largest residuals are
\begin{equation}
\max |F_{\rm ang}|=5.2\times10^{-14},
\qquad
\max |F_{\rm rad}|=2.51\times10^{-11},
\end{equation}
with maximum four-real-component norm
\begin{equation}
\max ||F||=2.51\times10^{-11}.
\end{equation}
All nonlinear solves return converged status at the stated tolerance. Repeating the survey at $(220,360)$ changes the complex frequency by at most
\begin{equation}
\max\left|\Delta(\omega/\mu)\right|=6.85\times10^{-9}.
\end{equation}
The normalization-controlled scan uses the same final cutoffs and has maximum joint residual $4.35\times10^{-11}$. Machine-readable tables and the audit software are available from the corresponding author.

\clearpage
\bibliographystyle{apsrev4-2}
\bibliography{KR_Dirac_references_scite_verified}

\end{document}